\documentclass[aps,prl,superscriptaddress,reprint,hyperref, amsmath]{revtex4-1}
\usepackage{graphicx, color}

\newcommand{\be}{\begin{equation}}
\newcommand{\ee}{\end{equation}}

\newcommand{\VL}{V_L}
\newcommand{\VR}{V_R}
\newcommand{\Vt}{V_t}

\newcommand{\tim}{t}
\newcommand{\detn}{\delta}
\newcommand{\dzero}{\detn_0}

\newcommand{\figa}{(a)}
\newcommand{\figb}{(b)}
\newcommand{\figc}{(c)}
\newcommand{\figd}{(d)}
\newcommand{\fige}{(e)}
\newcommand{\figf}{(f)}

\newcommand{\Eqo}{\varepsilon_{\textrm{q,0}}}
\newcommand{\Eqa}{\varepsilon_{\textrm{q,ad}}}

\newcommand{\ERWA}{\varepsilon_{\rm RWA}}
\newcommand{\quasiE}{\varepsilon}

\newcommand{\Ad}{A_{\rm d}}
\newcommand{\AdRWA}{A_{\rm d, RWA}}
\newcommand{\fd}{\nu_{\rm d}}

\newcommand{\fr}{\nu_{\rm r}}

\newcommand{\bra}[1]{\langle #1|}
\newcommand{\ket}[1]{|#1 \rangle}

\begin{document}


\title{Floquet spectroscopy of a strongly driven quantum dot charge qubit with a microwave resonator}


\author{J. V. Koski}
\affiliation{Department of Physics, ETH Zurich, CH-8093 Zurich, Switzerland}
\author{A. J. Landig}
\affiliation{Department of Physics, ETH Zurich, CH-8093 Zurich, Switzerland}
\author{A. P\'alyi}
\affiliation{Department of Physics, Budapest University of Technology and Economics, H-1111 Budapest, Hungary}
\affiliation{MTA-BME Exotic Quantum Phases "Momentum" Research Group, Budapest University of Technology and Economics, H-1111 Budapest, Hungary}
\author{P. Scarlino}
\affiliation{Department of Physics, ETH Zurich, CH-8093 Zurich, Switzerland}
\author{C. Reichl}
\affiliation{Department of Physics, ETH Zurich, CH-8093 Zurich, Switzerland}
\author{W. Wegscheider}
\affiliation{Department of Physics, ETH Zurich, CH-8093 Zurich, Switzerland}
\author{G. Burkard}
\affiliation{Department of Physics, University of Konstanz, D-78457 Konstanz, Germany}
\author{A. Wallraff}
\affiliation{Department of Physics, ETH Zurich, CH-8093 Zurich, Switzerland}
\author{K. Ensslin}
\affiliation{Department of Physics, ETH Zurich, CH-8093 Zurich, Switzerland}
\author{T. Ihn}
\affiliation{Department of Physics, ETH Zurich, CH-8093 Zurich, Switzerland}

\date{\today}


\begin{abstract}
We experimentally investigate a strongly driven GaAs double quantum dot charge qubit weakly coupled to a superconducting microwave resonator. The Floquet states emerging from strong driving are probed by tracing the qubit - resonator resonance condition.  This way we probe the resonance of a qubit that is driven in an adiabatic, a non-adiabatic, or an intermediate rate showing distinct quantum features of multi-photon processes and Landau-Zener-St\"uckelberg interference pattern.  Our resonant detection scheme enables the investigation of novel features when the drive frequency is comparable to the resonator frequency. Models based on adiabatic approximation, rotating wave approximation, and Floquet theory explain our experimental observations.
\end{abstract}

\maketitle


Applying a strong drive to a quantum two-level system (qubit) gives rise to intricate physics, such as ac Stark effect \cite{Autler1955}, multi-photon effects \cite{Saito2004}, and Landau-Zener-St\"uckelberg interference \cite{Shevchenko2010}, characterized by the emergence of Floquet states \cite{Shirley1965}. Many of these effects have practical benefit: if the qubit is coupled to a superconducting resonator, the ac Stark effect allows to calibrate the resonator photon number \cite{Schuster2005}. The Landau-Zener-St\"uckelberg interference pattern gives relevant information on qubit decoherence \cite{Wilson2007, Ribeiro2013, Forster2014}, and carefully considering the Floquet dynamics provides means to improve the fidelity of qubit operations \cite{Deng2016}. Strong driving dynamics have been investigated in various systems including superconducting qubits \cite{Oliver2005, Sillanpaa2006, Wilson2007, Izmalkov2008,Baur2009,  Tuorila2010, Oelsner2013, Deng2015, Chen2017} and quantum dot devices \cite{Petta2010, Stehlik2012, Forster2014,Forster2015,GonzalezZalba2016, Korkusinski2017, Bogan2017}. Recent experiments have shown that strongly driving a qubit that is coupled to a resonator can enhance the resonator transmission \cite{Oelsner2013, Neilinger2016, Stehlik2016, Wen2017}. Spectroscopy of the Floquet quasienergies of a strongly driven system has been demonstrated by following its time-evolution \cite{Fuchs2009, Deng2015}. An alternative proposal is to probe the driven qubit with a weakly coupled superconducting resonator \cite{Silveri2013, Kohler2017}. 


Here we report our experimental implementation of the proposal, where the resonance frequency of the resonator determines the probed Floquet quasienergy. Due to weak coupling, the resonator does not directly influence the qubit energetics. We perform a set of experiments, first with adiabatic driving with which we observe multi-photon processes and Landau-Zener-St\"uckelberg interference pattern similar as has been reported in other experiments \cite{Shevchenko2010, Stehlik2012, Forster2014}. In a second experiment we increase the driving rate from adiabatic to non-adiabatic, allowing us to probe the evolution of the Floquet quasienergy. In our third experiment, we apply a near-resonant and a near-half-resonant drive. There we observe a vanishing of the probe signal since the avoided crossing between the drive field photons and the qubit energy eliminates the states at the resonator energy. In this regime, all three energy scales present in our setup, i.e. those of the qubit, the resonator, and the drive photons, are relevant.


\begin{figure}[h!]
\includegraphics[width=\columnwidth]{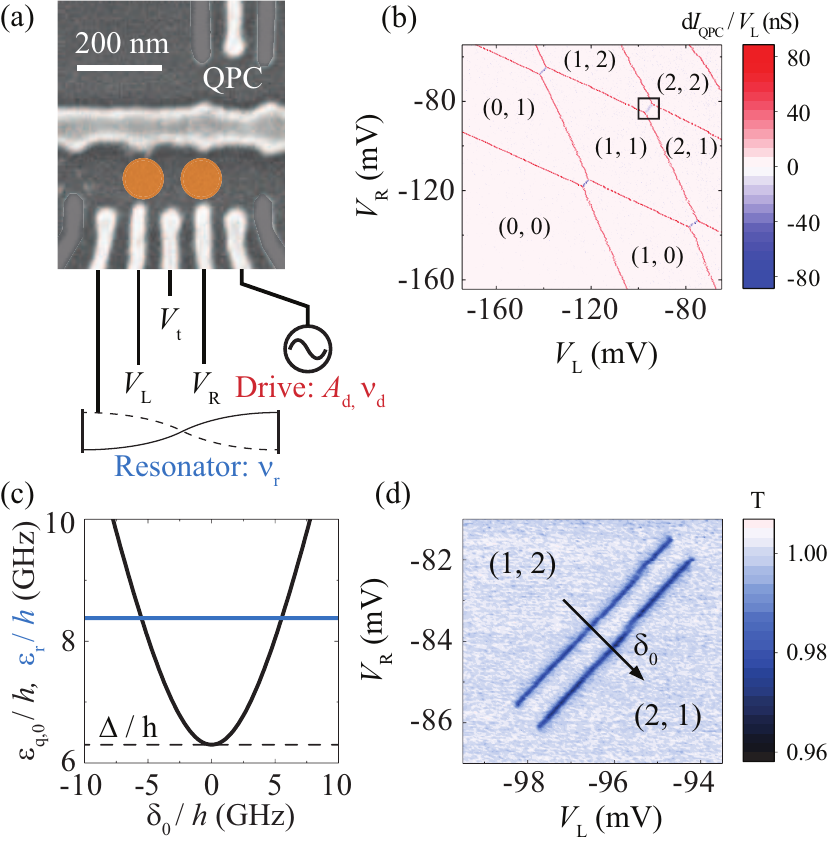}%
\caption{\figa~A scanning electron micrograph of the device. Two quantum dots are defined in the positions indicated by the orange dots. A superconducting resonator is connected to the leftmost gate, and the qubit is driven by applying a continuous tone to the rightmost gate. Unused gate electrodes are greyed out. \figb~Stability diagram of the double quantum dot measured by the QPC. The black rectangle marks the operation regime.  \figc~Charge qubit and resonator photon energy as a function of $\dzero$. The solid black line depicts the qubit energy $\Eqo = \sqrt{\Delta^2 + \dzero^2}$. The minimum qubit energy is determined by the interdot tunnel coupling $\Delta$ (black dashed line). The blue line shows the cavity photon energy $h \fr$. \figd~ Resonator transmission $T$ as a function of $\VL$ and $\VR$. The $\dzero$ axis is depicted with an arrow. 
\label{Fig1}} 
\end{figure}


Our qubit is formed in a double quantum dot (DQD), shown in Fig. \ref{Fig1}\figa. Au top gates define the DQD electrostatically in a two-dimensional electron gas hosted in a GaAs/AlGaAs heterostructure. The number of electrons in the DQD is controlled with plunger gate potentials $\VL$ and $\VR$ and monitored by a nearby quantum point contact (QPC). As indicated in the charge stability diagram Fig. \ref{Fig1}~\figb, the qubit is operated in the three electron regime where the relevant charge states are $\ket{L} = \ket{(2, 1)}$ and $\ket{R} = \ket{(1, 2)}$, with $(n, m)$ notation indicating $n$ electrons on the left and $m$ electrons on the right dot. In the $\ket{L}$ - $\ket{R}$ basis, the system Hamiltonian is $H_0 = \Delta \sigma_x / 2 + \dzero \sigma_z / 2$ where $\sigma_{x}$ and $\sigma_{z}$ are Pauli matrices, $\Delta$ is the interdot tunnel coupling, and $\dzero$ is the DC detuning energy between $\ket{L}$ and $\ket{R}$. We directly control $\Delta$ and $\dzero$ with $\Vt$, $\VL$, and $\VR$ shown in Fig. \ref{Fig1}~\figa. Diagonalization of $H_0$ determines the unperturbed qubit states with an energy separation of $\Eqo = \sqrt{\Delta^2 + \dzero^2}$ illustrated in Fig. \ref{Fig1}~\figc. 


The DQD is connected to a superconducting half-wavelength coplanar microwave resonator by extending the resonator voltage antinode to the drive gate indicated in Fig. \ref{Fig1}~(a), similarly as in previous work \cite{Frey2012}. This leads to a coupling of the charge qubit to the resonator electric field characterized by $H_\text{int} = g_0 \sigma_z(b^\dagger + b) / 2$, where $g_0$ is the zero-detuning ($\dzero = 0$) coupling strength, and $b^\dagger$ is the resonator photon creation operator. Our resonator has a resonance frequency $\fr = 8.32$~GHz and a linewidth $\kappa / 2\pi =$~110~MHz. By applying the methods in \cite{Frey2012}, we estimate $g_0 / 2\pi\approx$~30~MHz and a qubit decoherence of $\gamma_2 / 2\pi \approx 400$ MHz for our device. Since $g_0 \ll \gamma_2$, the resonator is weakly coupled to the qubit and allows weak probing without coherently influencing its states. When $\dzero$ satisfies the resonance condition $\Eqo = h\fr$, the qubit can absorb photons from the resonator. This is observed as a decrease in transmission when probing the resonator at frequency $\fr$. If $\Delta < h\fr$ as in Fig. \ref{Fig1}\figc, two qubit-photon resonances occur for $\dzero = \pm \sqrt{(h\fr)^2 - \Delta^2}$. By sweeping the gate potentials across the $(1, 2)$, $(2, 1)$ charge transition as shown in Fig. \ref{Fig1}\figd, the two resonances are observed.


We drive the qubit by applying a continuous microwave tone to the drive gate indicated in Fig. \ref{Fig1}\figa. This gives rise to a time-dependent detuning $\dzero \to \dzero + \Ad \cos(2\pi\fd \tim)$ and consequently a time-dependent Hamiltonian $H(t) = H_0 + \Ad \sigma_z \cos(2\pi\fd \tim) / 2$. The drive frequency $\fd$ has a significant effect on the qubit \cite{Oliver2005, Sillanpaa2006, Shevchenko2010, Forster2014}. As such, our experimental control parameters are the drive frequency $\fd$ and amplitude $\Ad$ as well as the qubit detuning $\dzero$ and tunnel coupling $\Delta$.


\begin{figure}[h!t]
\includegraphics[width=\columnwidth]{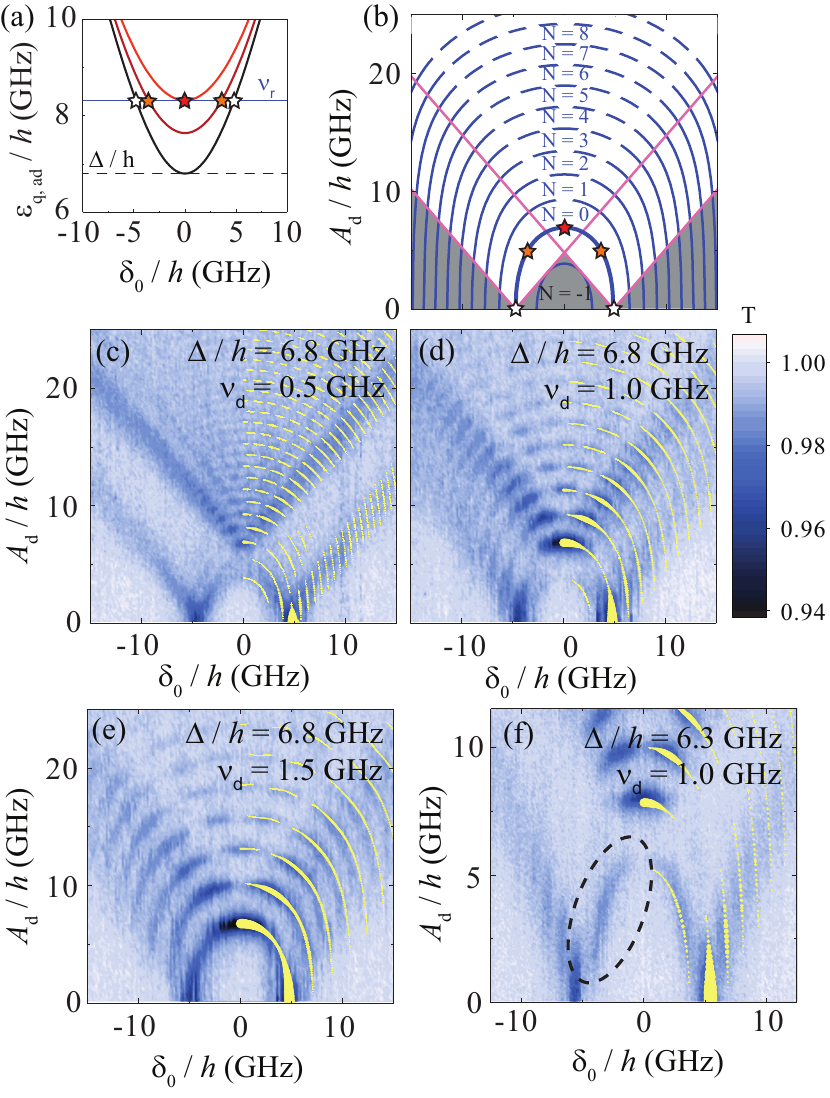}%
\caption{\figa~The effective energy of an adiabatically driven qubit as a function of $\dzero$ with drive amplitudes $\Ad/h =$ 0 GHz, 5 GHz, and $\sim$7 GHz in order of increasing energy (from black to red). (b) Illustation of the expected resonances as a function of $\dzero$ and $\Ad$ for $\fd = 1$ GHz and $\Delta / h = 6.8$ GHz. The resonance positions satisfying $\Eqa = h\fr + Nh\fd$ are shown as blue lines with labels displaying the number of photons absorbed from the drive field. The stars in panels (a) and (b) mark the corresponding resonance positions. The lines with unit slope emerging from the zero-$\Ad$ resonances define the grey regions where resonances are not visible. (c) - (f) show the measured transmission as a function of $\dzero$ and $\Ad$ for low $\fd$. The $N = -1$ resonance is indicated with a dashed ellipse in panel \figf. Yellow dots show the theoretically predicted resonance positions where one of the resonance conditions in Eq.~\eqref{eq:resonance} is fulfilled.
The radius of each yellow dot is proportional to the corresponding transition strength $\mathcal{M}$ in Eq.~\eqref{eq:visibility}. 
 \label{Fig2}}
\end{figure}


In our first experiment, we explore the low drive frequency regime $\fd \ll \Delta / h, \fr$. In this limit, the dynamics are approximately adiabatic and the effective qubit energy $\Eqa$ is given by the time-averaged qubit energy
\be \Eqa =\nu_d\int_0^{1/\fd}dt \sqrt{\Delta^2 + (\dzero + \Ad\cos(2\pi \fd t))^2}. \label{eq:driven_energy} \ee 
$\Eqa$ does not depend on $\fd$ and increases monotonically with increasing $\Ad$. As illustrated in Figs. \ref{Fig2}~\figa-\figb, increasing $\Ad$ will trigger the resonance condition $\Eqa = h\fr$ for a different $\dzero$ determined by the points where the energy of driven qubit and that of the resonator intersect. We also expect multi-photon resonances \cite{Supplementary} satisfying $h\fr = \Eqa - Nh\fd$, where $N$ is an integer. As sketched in Fig. \ref{Fig2}~\figb~for $\fd = 1.0$ GHz, this results in replica of the $N = 0$ resonance that are visible when $\Ad > |N|h\fd$.


We now measure the transmission for $\fd = 0.5$ GHz, $\fd = 1$ GHz, and $\fd = 1.5$ GHz, each with $\Delta / h = 6.8$ GHz, shown in Figs. \ref{Fig2}\figc-\figf. For zero drive amplitude, the qubit is resonant with the resonator at two detuning values $\dzero / h = \pm\sqrt{\fr^2 - (\Delta / h)^2} \simeq \pm 4.8$ GHz, see Fig. \ref{Fig2}\figa. Increasing $\Ad$ changes the resonance condition for $\dzero$ as illustrated in Fig. \ref{Fig2}\figb. With a sufficiently high drive frequency as in Fig. 2 \fige, the full $N=0$ resonance arcs are clearly discernible, while for $\fd = 0.5$ GHz and $\fd = 1.0$ GHz, the resonance visibility vanishes for the range $\Ad \gtrsim 4 h\fd, |\dzero| \gtrsim 2h\fd$. This feature is also captured by the predicted qubit visibility Eq. \eqref{eq:visibility}, discussed later. 


Multi-photon resonances matching $\Eqa = h\fr+ N h \fd$ emerge as $\Ad$ is increased. Qualitatively, these processes take place by the qubit absorbing a single photon from the resonator, and $N$ photons from the drive field. For these $N$ + 1 photon processes, there are $N$ gaps in the resonance arcs, symmetrically distributed around $\dzero = 0$ as sketched in Fig. \ref{Fig2} \figb. This effect is reminiscent of Landau-Zener-St\"uckelberg interference \cite{Shevchenko2010}. Aside the interference pattern, the multi-photon resonances are qualitatively similar to that of the $N=0$ resonance, including the vanishing visibility regimes with $\fd = 0.5$ GHz and $\fd = 1.0$ GHz. We further perform a measurement with $\fd  =1.0$ GHz and $\Delta / h = 6.3$ GHz shown in Fig. \ref{Fig2}~\figf~to find a clear signature of a resonance corresponding to $N = -1$. This implies a process where the resonator photon is absorbed  by the qubit and as a single photon by the drive field. 


\begin{figure} [h!t]
\includegraphics[width=\columnwidth]{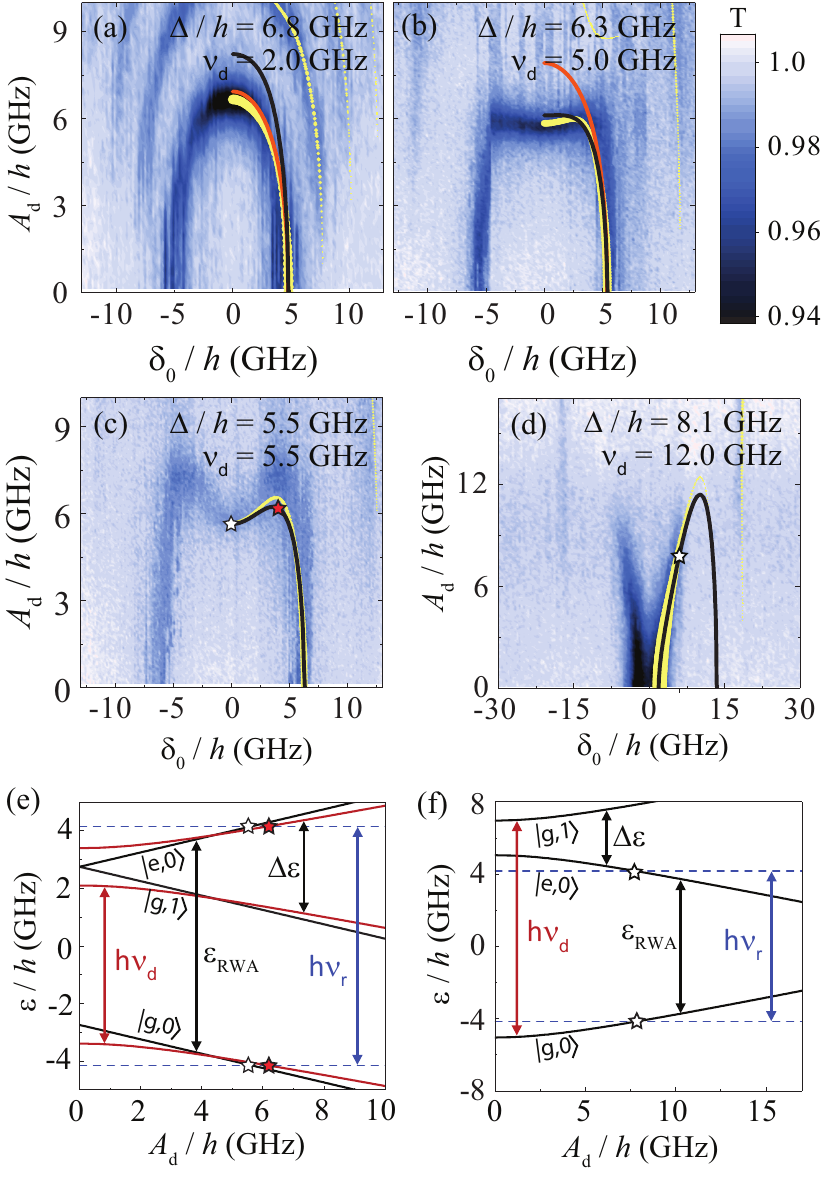}%
\caption{\figa-\figd~Measured transmission for increasing drive frequency. Yellow dots show the resonance conditions in Eq.~\eqref{eq:resonance}, and their radius proportional to the corresponding transition strength $\mathcal{M}$ in Eq.~\eqref{eq:visibility}. The black lines show the RWA resonance conditions by Eq. \eqref{eq:RWACondition}, and orange lines (in panels \figa~and \figb) show the adiabatic resonance conditions $\Eqa = h\fr$ with Eq. \eqref{eq:driven_energy}. \fige~The energy of the states $\ket{g, 0}$, $\ket{g, 1}$, and $\ket{e, 0}$ by RWA as a function of drive amplitude for parameters in \figc. Black lines correspond to $\dzero/h = 0$ GHz, and red lines to $\dzero/h = 4$ GHz. The splitting $\Delta \varepsilon = \sqrt{\left(\Ad\sin(\theta)/2\right)^2 + \left(\Eqo - h\fd\right)^2}$ between $\ket{g, 1}$ and $\ket{e, 0}$ is determined by Eq. \eqref{eq:RWA_Hamiltonian}. The resonance condition for $\ket{g, 0}$ and $\ket{e, 0}$ with $h\fr$ (dashed blue line) is marked by stars. \figf~Corresponding plot for parameters in \figd~and $\dzero/h = 6$ GHz.   \label{Fig3}}
\end{figure}


To describe our data more precisely, we use the quantum-electrodynamics interpretation of Floquet theory \cite{Shirley1965}. In this language, the coupled system of the charge qubit and its driving field is described by the Hamiltonian \cite{Supplementary} $H_\text{F} = H_0 + h \nu_\text{d} \hat{m} + A_\text{d} \sigma_z (\hat{m}_- + \hat{m}_+) / 4$,
where the basis is $\ket{L,m}$, $\ket{R,m}$, with $m$ corresponding to the number of photons in the drive field, $\hat{m} = \sum_m m \ket{m} \bra{m}$ is the drive field photon number operator, and $\hat{m}_+ = \hat{m}^\dag_- = \sum_m \ket{m+1} \bra{m}$. We calculate numerically two Floquet eigenstates $\ket{\pm} = \sum_{m} c^\pm_{L,m} \ket{L,m} + c^\pm_{R,m} \ket{R,m}$ with corresponding quasienergies $\quasiE_\pm$ that satisfy $- h \fd/2 < \quasiE_- \leq \quasiE_+ \leq h \fd/2$ \cite{Supplementary}. All other eigenstate-quasienergy pairs of $H_F$ are shifted replicas of these two, satisfying $\ket{\pm,n} = \left(\hat{m}_+\right)^n \ket{\pm}$ and $\varepsilon_{\pm,n} = \varepsilon_{\pm} + n h \nu_\text{d}$. The driven qubit can absorb a resonator photon if
\be
\label{eq:resonance}
h\nu_\text{r} = \varepsilon_{\pm,n} - \varepsilon_{\mp},
\ee
where $n$ is a non-negative integer. We locate numerically the resonance positions ($\dzero$, $\Ad$) that satisfy one of the resonance conditions in Eq. \eqref{eq:resonance}, and show them  in the transmission plots in Figs. 2 (c)-(f), Fig. 3, and Fig. 4. Overall, the resonance positions match accurately with the transmission minima seen in the experiment. In the spirit of Fermi's Golden Rule, the strength $\mathcal{M}$ of the corresponding transition is characterized by the squared matrix element of the qubit-probe coupling Hamiltonian $H_\text{int} \propto \sigma_z$ between the initial and final states \cite{Silveri2013}, 
\be \label{eq:visibility} \mathcal M = \left|\bra{\pm,n} \sigma_z \ket{\mp} \right|^2. \ee


In our second experiment, we move to the regime where $\fd$ is comparable to our other frequency scales $\Delta / h$ and $\fr$ and concentrate on the $N = 0$ resonance. Figures \ref{Fig3} \figa-\figd~show the measured transmission as a function of $\dzero$ and $\Ad$, displaying the evolution of the resonance pattern as the configuration transitions from adiabatic to non-adiabatic regime. While Fig. \ref{Fig3}\figa~shows the arc-shaped resonance characterized by the adiabatic approximation Eq. \eqref{eq:driven_energy}, for a higher drive frequency shown in Figs. \ref{Fig3}\figb-\figd, such approximation is no longer sufficient. Here, the signatures of non-adiabaticity are the non-monotonicity of the resonance condition in $\Ad$ as a function of $\dzero$ in Figs. \ref{Fig3}~\figb-\figc, and branching of $\Ad$ in Fig. \ref{Fig3}~\figd.


To interpret the high frequency drive data, we use a rotating wave approximation (RWA). We change from $\ket{L}$, $\ket{R}$ basis to the qubit eigenstate basis with $\Ad = 0$, i.e. $\ket{g} = -\sin(\theta/2) \ket{L} + \cos(\theta / 2)\ket{R}$, $\ket{e} = \cos(\theta/2) \ket{L} + \sin(\theta/2)\ket{R}$, where the mixing angle $\theta$ is given by $\cos(\theta) = \dzero / \Eqo$. We then limit our basis to two interacting states close in energy, such as $\ket{e, m = 0}$ and $\ket{g, m = 1}$, for which we can write the Hamiltonian as
\be H_{\rm RWA} = \frac{\Eqo}{2}\sigma_z + \frac{A_d}{4}\sin(\theta)(\sigma_+ {\hat m}_- + \sigma_- {\hat m}_+) + h\fd \hat m, \label{eq:RWA_Hamiltonian} \ee
where $\sigma_+ = \sigma_-^\dagger = \frac{1}{2}\left(\sigma_x + i\sigma_y\right)$ is the qubit raising operator. As indicated in Fig. \ref{Fig3}\fige, the energy of $\ket{g, m = 1}$ is offset from the energy of $\ket{g, m = 0}$ by $h\fd$ and we determine \cite{Supplementary} the effective qubit energy from Eq. \eqref{eq:RWA_Hamiltonian} as $\ERWA = h\fd \pm \sqrt{\left(\Ad\sin(\theta)/2\right)^2 + \left(\Eqo - h\fd\right)^2}$. The resonance condition $\ERWA = h\fr$ for $\Ad$ is then
\be \AdRWA = \frac{2}{\Delta} \Eqo\sqrt{(h\fr - h\fd)^2 - (\Eqo - h\fd)^2}. \label{eq:RWACondition} \ee


The data shown in Figs. \ref{Fig3}\figb~and~\figc~ lie in the regime where we find a non-monotonic resonance condition as predicted by Eq. \eqref{eq:RWACondition}. Qualitatively, the dip in $\AdRWA$ at $\dzero = 0$ is due to a maximum in the qubit - drive photon coupling strength, which is proportional to $\Ad\sin(\theta) = \Ad\Delta / \sqrt{\Delta^2 + \dzero^2}$ as in Eq. \eqref{eq:RWA_Hamiltonian}, leading to faster change in $\ERWA$ with increasing $\Ad$ as shown in Fig. \ref{Fig3}\fige. Figure \ref{Fig3}\figd~shows transmission data for $\fd = 12.0$ GHz and $\Delta/h = 8.1$ GHz. In this regime with $\fd > \fr$, we observe that the resonance condition for $\Ad$ increases with increasing $\dzero$. This can be understood from Fig. \ref{Fig3}\figf, showing that when $h\fr < \Eqo < h\fd$, increasing $\Ad$ lowers the qubit energy and brings the qubit on resonance with the resonator. We find that both the RWA result in Eq. \eqref{eq:RWACondition} and the transition strength by Eq. \eqref{eq:visibility} give a good prediction for the resonance locations for datasets shown in Figs. \ref{Fig3}\figb-\figd. We also note that in each measurement shown in Fig. \ref{Fig3}, the probe signal is continuously visible.


\begin{figure} [h!t]
\includegraphics[width=\columnwidth]{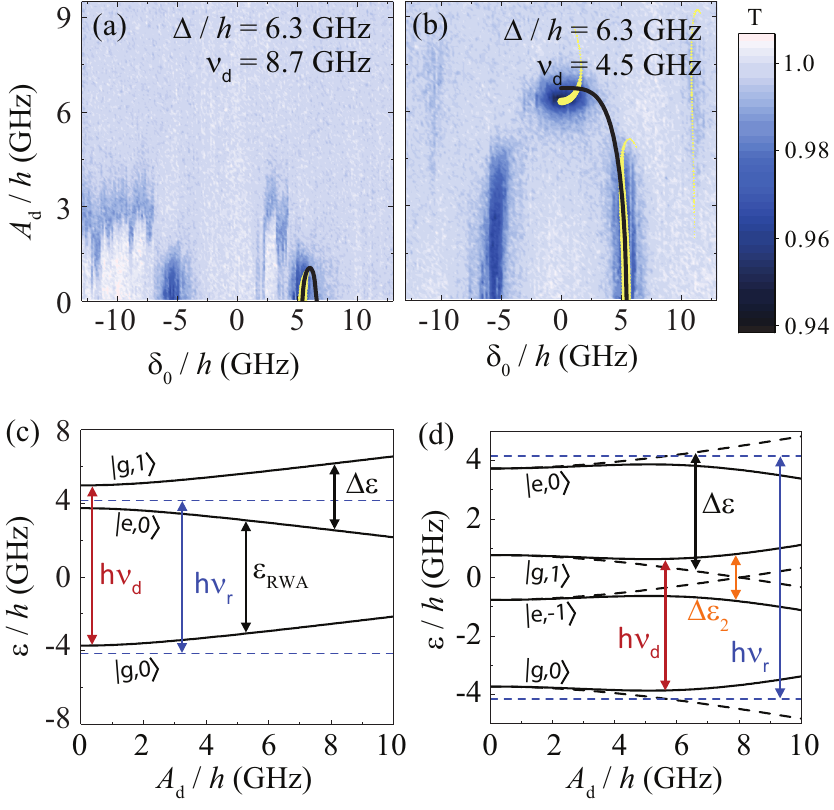}
\caption{Transmission with \figa~near-resonant driving $\fd \simeq \fr$ and \figb~half-resonant driving $\fd \simeq \fr / 2$ as a function of detuning and drive amplitude. Yellow dots show the resonance conditions in Eq.~\eqref{eq:resonance}, and their radius is proportional to the transition strength in Eq.~\eqref{eq:visibility}. The black lines show the RWA resonance conditions by Eq. \eqref{eq:RWACondition}. \figc~ Energy levels  with $\dzero / h = 4$ GHz. In panel \figd, the RWA results are shown as dashed lines. Second order effects \cite{Supplementary} form an avoided crossing $\Delta \epsilon_2 \propto \Ad^2 \cos(\theta)\sin(\theta)$ between the RWA states. \label{Fig4}}
\end{figure}


In our third experiment, $\fd$ is near-resonant or near-half-resonant with $\fr$. Figure \ref{Fig4}~\figa~shows the transmission for $\fd = 8.7$ GHz, which is close to the resonator frequency $\fr \simeq 8.32$ GHz. We observe that the transmission signal vanishes as $\Ad$ is increased, an effect that can be understood from the RWA Hamiltonian Eq. \eqref{eq:RWA_Hamiltonian}. As illustrated in Fig. \ref{Fig4}\figc, when $\Eqo \simeq h\fd$, the qubit energy changes rapidly from $h\fd$ with increasing $\Ad$ due to $\ket{e, 0} - \ket{g, 1}$ hybridization. Therefore, if $\fd = \fr$, the driven qubit cannot have an energy exactly matching $h\fr$. Figure \ref{Fig4}~\figb~shows transmission for $\fd = 4.5$ GHz which close to half $\fr$. In this regime, the RWA result is no longer sufficient to characterize the observed resonances. The special case of half-harmonic driving is discussed in \cite{Romhanyi2015}: here, the qubit is influenced by second order photon processes where the hybridization gap arises when $2\fd \simeq \ERWA / h$, as illustrated in Fig. \ref{Fig4}\figd. However, second order coupling scales with $\Ad^2\cos(\theta)\sin(\theta)$ \cite{Supplementary}, which tends to zero when $\dzero \rightarrow 0$. This gives rise to the small range in $\dzero$ where the qubit remains visible.


In conclusion, we have comprehensively investigated Floquet energy spectra of a strongly driven charge qubit with a weakly coupled microwave resonator as a function of qubit detuning and drive amplitude over a large range of drive frequencies. In contrast to earlier experiments studying strongly driven systems, we have explored the regime where the qubit can be brought on resonance with the resonator either by detuning, or by increasing drive amplitude. This feature allows to extract the Floquet quasienergy spectrum of a strongly driven charge qubit. The experiment could be extended towards measuring the adiabatic phases of a doubly-driven qubit \cite{Kohler2017}.

\acknowledgments
We thank Benedikt Kratochwil for a critical discussion. This work was supported by the Swiss National Science Foundation through the National Center of Competence in Research (NCCR) Quantum Science and Technology. 
AP was supported by the National Research Development and Innovation Office of Hungary within the Quantum Technology National Excellence Program (Project No. 2017-1.2.1-NKP-2017-00001) and Grants 105149 and 124723, and by the New National Excellence Program of the Ministry of Human Capacities. GB ackowledges funding from DFG within SFB 767.

\bibliography{references}

\end{document}